\documentclass[useAMS,fleqn,usenatbib]{mnras}

\usepackage{color,amsmath,physymb}
\usepackage[pdftex]{graphicx}

\newcommand{\rmd}{\mathrm{d}}
\newcommand*{\hm}{\Gamma}

\title[Search for wormhole candidates in active galactic nuclei...]{Search for wormhole candidates in active galactic nuclei: radiation from colliding accreting flows}

\author[M.Yu. Piotrovich et al.]{
M.Yu. Piotrovich \thanks{E-mail: mpiotrovich@mail.ru},
S.V. Krasnikov \thanks{E-mail: krasnikov.xxi@gmail.com},
S.D. Buliga
T.M. Natsvlishvili,
\\
Central Astronomical Observatory at Pulkovo, 196140, Saint-Petersburg, Russia}

\date{Accepted for publication in MNRAS.}

\pubyear{2020}

\begin{document}

\label{firstpage}
\pagerange{\pageref{firstpage}--\pageref{lastpage}}
\maketitle

\begin{abstract}
The underlying hypothesis of this work is that the active galactic nuclei (AGNs) are wormhole mouths rather than supermassive black holes (SMBHs). Under some---quite general---assumptions such wormholes may emit gamma radiation as a result of a collision of accreting flows inside the wormholes. This radiation has a distinctive spectrum much different from those of jets or accretion disks of AGNs. An observation of such radiation would serve as evidence of the existence of wormholes.
\end{abstract}


\section{Introduction} %1

\begin{figure}
	\label{fig01}
  \centering
  \includegraphics[bb= 130 15 460 465, clip, width=0.7\columnwidth]{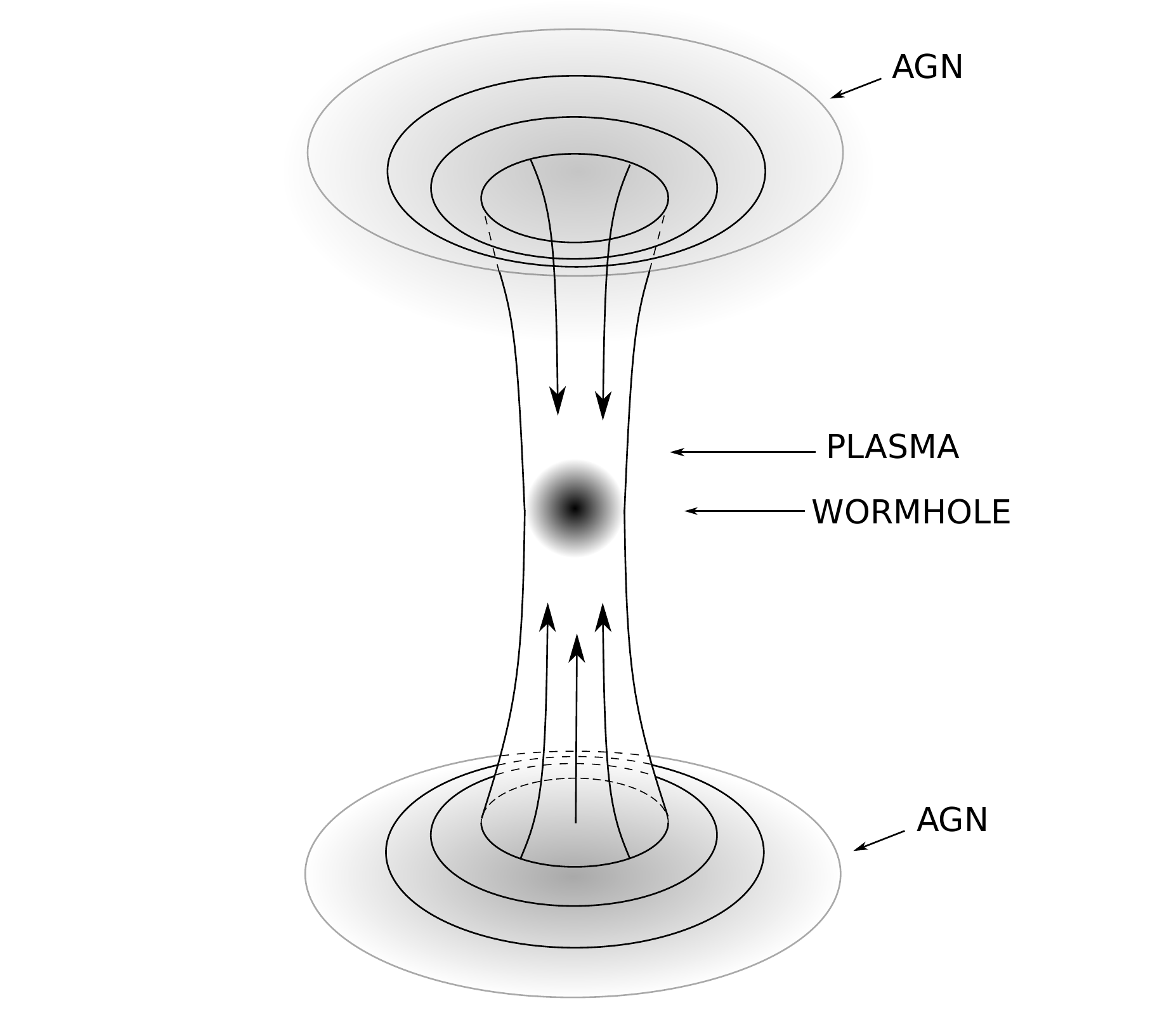}
  \caption{Traversable wormhole connecting two AGNs.}
\end{figure}

According to an interesting recent hypothesis AGNs are not SMBHs, but rather mouths of primordial macroscopic wormholes \citep{kardashev07,bambi13,li14,zhou16}. In this paper we show that the validity of  such a  hypothesis may have observable consequences.

\section{Model and basic equations}

Consider a traversable wormhole connecting two galactic centers (see Fig.\ref{fig01}). To make the situation as simple as possible assume that the wormhole is spherically symmetric, static and empty everywhere beyond a spherical layer $L$. Assume further that the gravitational force acting on a radially falling test particle in any point of that wormhole is directed towards the throat.

An example of a wormhole satisfying all these requirements is a ``Schwarzschild-like wormhole''

\begin{equation}
 \label{eq:SchWh}
 \rmd s^2 =-e^{2 h(l)}c^2\rmd t^2 + \rmd l^2 + r^2(l) (\rmd \theta^2 +\sin^2  \theta \,\rmd \phi^2),
 \, l\in \mathrm{I\!R},
\end{equation}

\noindent  studied in its capacity as a particle accelerator in \citep{krasnikov18}.  $c$ in \eqref{eq:SchWh} is the speed of light, $l$ is (up to the sign) the proper radial distance from the wormhole's throat \citep{morris88}, and  $r(l)$ and $h(l)$ are smooth  even functions  monotone increasing at positive $l$   (almost all these conditions may be weakened, but we consistently keep avoiding unnecessary complications)  and obeying the following relations

\begin{equation}
 \rmd r = \sqrt{ 1-1/x}\,\rmd l, \qquad x\equiv r/r(0),
\end{equation}

\begin{equation}
 \label{eq: phi}
 h(x)\evalat{}{r>r_*}{}=\tfrac 12\ln( 1-1/x),
\end{equation}

\noindent where $r_*$ is some constant   greater than $r(0)$. We choose this constant  to be close to the radius of the wormhole's throat

\begin{equation}
 \label{eq:r*}
 2r(0) >r_* >  r(0),
\end{equation}

\noindent i.~e. the wormhole is required to be short (again, this condition might be dropped and $r_*$ be  made a free parameter of the model, but we don't need such generality,  while the hypothesis \eqref{eq:r*} will be helpful for the subsequent calculations).  Thus, for all $r > r_*$  the metric (\ref{eq:SchWh}) is that of Schwarzschild with mass

\begin{equation}
 \label{eq:gravr}
 M\equiv \frac{c^2}{2G}r(0)\approx\frac{c^2}{2G}r_*.
\end{equation}

\noindent At smaller $r$, the shape of $h (l)$  [\emph{not of $r(l)$}] differs from its Schwarzschild counterpart $h_\text{Sch}$, i.~e. from $\tfrac 12\ln( 1-1/x) $, which means that this region  is non-empty (it is, in  fact, the layer $L$ mentioned above). As shown in \citet{morris88}, if the Einstein equations hold, the matter filling $L$ unavoidably is \emph{exotic}, i.~e. its energy density as measured by  some observers in some points (including the points of the sphere $l=0$) is negative. There are several proposals in the literature on what could serve as exotic matter, see, e.~g.,\ \citet{lobo17} and \citet{visser95} for a review, but in this paper we  shall not consider them (for lack of need: as long as  the matter supporting the wormhole does not interact electromagnetically, its nature is irrelevant). Two remarks, however, are in order:\\
1). The idea that the Einstein equations require exotic matter to have $T_ {tt}<0$ and a Schwarzschild-like wormhole  to have $M<0$  is oversimplification. In particular, the wormhole \eqref{eq:SchWh} has $T_ {tt}=0$ and $M>0$ \citep{krasnikov18}. This can be seen from eq.~(12) in \citet{morris88} given that the function $b$ appearing there is constant in our case.\\
2). As ``normal'' matter---with the  stress-energy tensor $t_{\mu\nu}$  accretes to a wormhole (in such a manner that $t_{\mu\nu}$ remains diagonal in the coordinate basis), the matter inside the latter  presumably becomes ``less and less exotic'' (assuming that the stress-energy tensor near the throat is the sum $T_ {\mu\nu}+ t_{\mu\nu}$, which is justified by our condition that the two kinds of matter do not directly interact). Consequently, in some time $\mathcal{T}_\text{stab}$ the components of  $t_{\mu\nu}$ may cease to be negligible and the wormhole may lose stability and collapse. Unfortunately, for the reasons discussed in the previous item, it is hard to find an estimate on $\mathcal{T}_\text{stab}$ valid for all types of wormhole.

The most important difference between  $h_\text{Sch}$ and  $h$ is that the latter, in contrast to the former, is bounded. That is to say the spacetime \eqref{eq:SchWh} is a wormhole rather than a black hole. It has no horizon and is therefore traversable \citep{morris88}. This fact makes it possible to use a Schwarzschild-like wormhole as super accelerator. Indeed, the matter attracted by mouths of such a wormhole will form two fluxes colliding at the throat. To analyze  this process, consider, first, a particle of mass $\mu$ that falls freely from infinity (with zero initial velocity) and in the throat of the wormhole \eqref{eq:SchWh} hits head-on exactly the same oncoming particle. The energy of \emph{this} collision (in the center-of-mass system) is, see eq.~(12) in  \citep{krasnikov18},

\begin{equation}
 \label{eq:en}
 \begin{split}
 & E_{\rm c.m.}  = 2 \mu c^2\hm [1+\mathcal{O}(\hm^{-2})],\\
 & \hm\equiv c \max \Bigl( |g_{tt}|^{-1/2}\Bigr) =e^{-h _\text{min}},
 \end{split}
\end{equation}

\noindent where $g_{tt}$ is the corresponding component of the metric \eqref{eq:SchWh}. Note that the function $h$ attains its minimal value $h _\text{min}$ at the throat $l=0$. So, $h(0)$ is the only wormhole's characteristic on which $E_{\rm c.m.}$ depends. Though $h(0)$ is fixed for any given wormhole, one may call the Schwarzschild-like wormhole a super accelerator, because that fixed value can be   arbitrarily high.

As a result of the pair collisions discussed above,  an  observer watching the wormhole  will see two spheres (these are the spheres with radii $r_*$ bounding $L$) of ultra relativistic plasma. These spheres radiate with a distinctive spectrum, see bellow, much different from those of jets or accretion disks, which makes it possible to identify the wormhole.

\section{Estimates}

\subsection{Assumptions}

The phenomenon in discussion is rather complex and to explore it we will make as much simplifying assumptions as possible. Specifically, we consider a pair of AGNs with the masses $M \approx 5 \times 10^8 M_\odot$ connected by a Schwarzschild-like wormhole \eqref{eq:SchWh}.  We assume that

\begin{enumerate}
 \item the galaxies in question are Narrow-line Seyfert 1 Galaxies (NLS1) \citep{paliya19,kynoch19,madejski16,baghmanyan18},  which have a fairly weak gamma radiation, so our supposed WH gamma radiation will be less diluted;
 \item the plasma layer is heated (only) by the collision of the fluxes;
 \item the magnetic field, the interaction of the infalling matter with the outcoming radiation, and the backreaction of the spheres on the metric of the wormhole are negligible.
\end{enumerate}

 Now we can roughly estimate plasma temperature and spectrum.

\subsection{Plasma parameters}

The characteristic accretion rates of NLS1 lie \citep{bian03,collin04} within the range of

\begin{equation}
 \label{eq:rate}
 \dot{m} \approx 0.1 - 1 M_{\odot} / \textrm{year}.
\end{equation}

According to estimates of duty cycles of AGNs of this type \citep{shankar13,aversa15,tucci17}, this accretion regime should exist for at least $10^8$ years. So, the mass of the plasma spherical layer is at least

\begin{equation}
 \label{eq:m}
 m \approx 10^8 M_{\odot},
\end{equation}

\noindent where the layer's mass $m$ is understood as the integral of density rather than a characteristic of the wormhole's metric.

Now let us estimate the   proton concentration in the accreting plasma. To that end assume (following \citep{bian03,collin04}) that the region  external to  a sphere  $r=const\gg r_*$ is left by $M_{\odot}$ of matter in a year. Then (since  the matter does not accumulate between that sphere and the sphere  $r=r_*$) the latter will be left by exactly the same amount of plasma every year of the \emph{coordinate} time $t$. This plasma moving with (almost) the speed of light,  will fill the layer of proper (with respect to an observer whose $r$-, $\phi$-, and $\theta$- coordinates are constant) volume $V_\text{prop} =4\pi r_*^2 \cdot \sqrt{-g_{tt}(r_*)}\cdot1\,$ly. So, the proton concentration is

\begin{align}
  &\nonumber n_{acc} = N/V_\text{prop}\approx
   M_{\odot}\hm/ ( m_p 4\pi r_*^2 \cdot1\,\mathrm{ly})\\ \nonumber & \approx
  \frac{\hm}{(M/M_{\odot})^2}\cdot\frac{M_{\odot}}{ m_p} \frac{1}{\pi \, (10^{11}\,\mathrm{cm}^2)\cdot (1\,\mathrm{ly} )}
       \\ &\nonumber\approx
       \frac{\hm}{  (M/M_{\odot} )^2} \cdot\frac{2\cdot 10^{30}}{ 1.7 \cdot 10^{-27}} \frac{10^{-18}\,\mathrm{cm}^{-1} }{\pi \, (10^{11}\,\mathrm{cm}^2) }
  \\&\approx  \frac{\hm}{  (M/M_{\odot} )^2}\times 4\cdot 10^{27}\,\mathrm{cm}^{-3}
\label{eq:n acc}
\end{align}

\noindent where  $m_p$ is the  proton mass and $N$ is the number of protons in $V_\text{prop}$.

If we take an AGN mass value of $M \approx 5 \times 10^8 M_\odot$ and $\hm \approx 10$, then $n_{acc} \approx 1.6\cdot 10^{11}\, cm^{-3}$.

\subsection{Temperature and spectrum}

\begin{figure}
	\label{fig02}
  \centering
  \includegraphics[bb= 50 40 725 540, clip, width=\columnwidth]{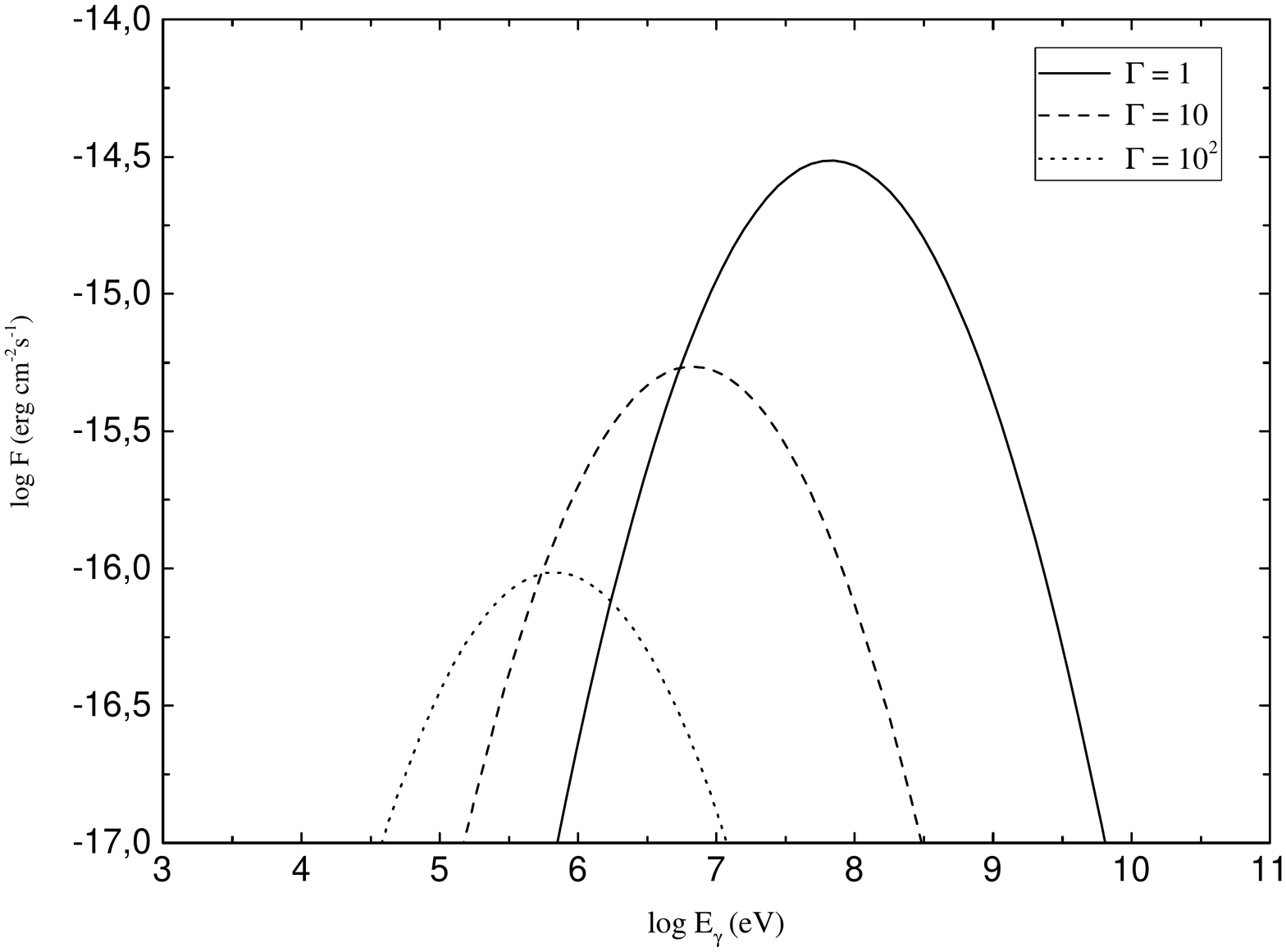}
  \caption{The spectrum of the ultrarelativistic plasma (for $V/(4\pi D^2) \approx 0.1\, cm$ (\ref{eq:flux})) for Minkowski case ($\hm = 1$) and for wormhole case ($\hm = 10,\, 100$).}
\end{figure}

By Eq.~(\ref{eq:en}) the temperature $T$ of the plasma cloud is approximately

\begin{equation}
 \label{eq:est T}
 T \approx \hm  m_p c^2 / k \approx 10^{13} \hm \,K,
\end{equation}

\noindent where $m_p$ is the proton mass and $k$ is the Boltzmann constant.

The plasma with such a temperature and proton concentration was considered, for example, in \citet{marscher80}. So in what follows we will use their results.

\subsubsection{Spectral peak}

At such high temperatures in the plasma the production of $\pi^0$ mesons begins \citep{kolykhalov79,marscher80,dermer86}. $\pi^0$ mesons decay into two photons with energies $\sim 68$ MeV. For $T = 2.5\times 10^{13}$K and $T = 10^{14}$K the gamma-ray production spectra of plasma for this radiation mechanism (in Minkowski space) were found in \citet{marscher80}. The latter spectrum has a peak value of $Q_\text{max} \approx 10^{-13.4} erg\, cm^{-3} s^{-1}$  at about 68$\,$MeV, where $Q$ is the gamma-ray production value. To find the spectrum of the wormhole (as measured by an observer resting out of the wormhole) we extrapolate Marscher's result (Fig.3 from \citet{marscher80}) and assume that the peak value in the spectrum is reached at $\sim 68 \hm^{-1} \,$MeV  (the factor $\hm^{-1}$ is due to the red shift experienced by a photon on its way from the throat to infinity) and equal to $Q_\text{max}(T) \approx Q_\text{max}(T = 10^{14}K) (T / 10^{14}K)^{0.25} / \hm$.

\subsubsection{Observability}

In order to evaluate possible observed flux value of plasma cloud we will use the relation \citep{marscher80}:

\begin{equation}
 \label{eq:flux}
 F(E_\gamma) = \frac{V}{4\pi D^2} Q(E_\gamma),
\end{equation}

\noindent where $E_\gamma$ is the photon energy, $F$ is the flux value, $V$ is the total volume of the plasma cloud and $D$ is the distance to the observer.

In order to obtain peak flux value comparable in order of magnitude to observed values for NLS1 ($\sim 10^{-13} erg\, cm^{-2} s^{-1}$) using Eq.(\ref{eq:flux}), we have to assume that $V / (4 \pi D^2) \geq 0.1$cm. Characteristic distance for observed AGNs of this type is about $10^9$  light-years or $\sim 10^{27}$cm, so the volume of the radiating plasma must be at least $10^{54} cm^3$. In \citep{marscher80} the plasma is optically thin, so we consider radiation only from outer layers of plasma cloud with a volume of 10\% of the total volume of cloud. Thus full cloud volume will be $\sim 10^{55} cm^3$. Taking into account \eqref{eq:n acc} we obtain the mass of the plasma cloud

\begin{equation}
 \label{eq;mM}
 m = n_{acc}m_pV \approx 2.4 \times 10^{38} kg \sim 10^8 M_{\odot}
\end{equation}

\noindent Comparing this with characteristic mass of accreted matter for NLSy1 $10^8 M_{\odot}$  \eqref{eq:m}, we see that the NLSy1 observed accretion rate $0.1 - 1 M_{\odot} / year$  \eqref{eq:rate} is sufficiently high for WH radiation to be observable.

The resulting spectrum for $V/(4\pi D^2) \approx 0.1\, cm$ (\ref{eq:flux}) looks (qualitatively) as shown in Fig.2.

\section{Conclusions}

We can conclude that a peak at about $68 \hm^{-1} \,$MeV should appear on the AGN spectrum. It should be emphasized that this peak in the spectrum can be quite high ($\sim 10^{-13} erg\, cm^{-2} s^{-1}$), comparable to flux value of jet radiation of NLS1.

\section*{Acknowledgements}

S.K. is grateful to RFBR for financial support under grant No.~18-02-00461 "Rotating black holes as the sources of particles with high energy".

\section*{Data availability}

The data underlying this article are available in the article.

\bibliographystyle{mnras}
\bibliography{mybibfile}

\bsp
\label{lastpage}
 \end{document}